\documentclass[prb,twocolumn,aps,showpacs,preprintnumbers]{revtex4}
\usepackage{graphicx}
\usepackage{dcolumn}
\usepackage{bm}

\begin{document}

\preprint{Draft}

\title{Spin Hall effects in diffusive normal metals.}

\author{R.V. Shchelushkin and Arne Brataas}
\affiliation{Department of Physics, Norwegian University of Science 
and Technology, N-7491 Trondheim, Norway}
\date{\today}

\begin{abstract}
We consider spin and charge flow in normal metals. We employ the Keldysh formalism to find transport equations in the presence of  
spin-orbit interaction, interaction with magnetic impurities, and non-magnetic impurity scattering. Using the quasiclassical approximation, we derive diffusion equations which include contributions from skew scattering, side-jump scattering and the anomalous spin-orbit induced velocity. We compute the magnitude of various spin Hall effects in experimental relevant geometries and discuss when the different scattering mechanisms are important.    
\end{abstract}


\maketitle

\section{INTRODUCTION}

Spin flow in nanostructures has recently attracted considerable interest in the scientific community\cite{GMROptics1,GMROptics2,GMROptics3}. The vision of magnetoelectronics and spintronics is to inject, manipulate and detect spins in nanostructures which can give new functionality in electronic devices. The spin flow can be controlled by \textit{e.g.} external electric or magnetic fields.

A Hall voltage builds up perpendicularly to the current flow under an applied magnetic field in normal metals due to the Lorentz force. The Hall voltage increases with applied magnetic field. Magnetoelectronic circuits are often realized by using ferromagnets that can spin polarize the current flow. In ferromagnets, there is an anomalous Hall voltage proportional to the magnetization, \textit{e.g.} a transverse charge potential, even in the absence of an applied magnetic field. The anomalous Hall effect is caused by the spin-orbit interaction, which correlates the momentum of the electron with its spin. This causes an dependence of the electron flow with the relative angle between its direction and the non-zero magnetic order parameter in ferromagnets.\cite{bruno,hirsh,dyak} In ferromagnetic metals, spin-orbit interaction is also a source of crystalline magnetic anisotropy energies since the spin-orbit interaction couples the magnetization with the crystal structure. 

\begin{figure}[ht]
\includegraphics[angle=0,width=7cm]{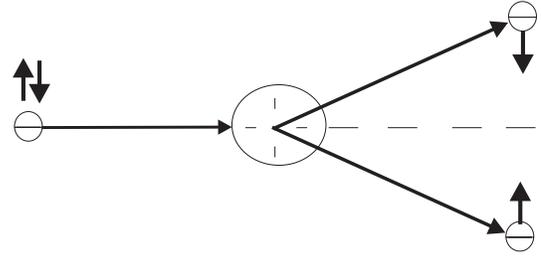}
\caption{Schematic picture of the skew scattering mechanism. An incident electron with spin up (down) scatters preferrably with a postive (negative) angle.}
\label{fig:skew}
\end{figure}

\begin{figure}[ht]
\includegraphics[angle=0,width=7cm]{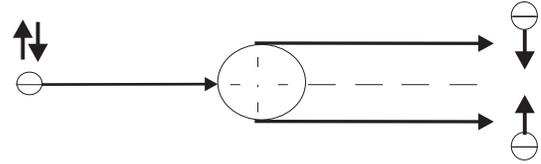}
\caption{Schematic picture of the side-jump mechanism. The trajectory of the outgoing electrons is shifted to the upper (lower) side the scattering center at large distances for spin up (down) states. }
\label{fig:sidejump}
\end{figure}

The scattering mechanisms responsible for the anomalous Hall effect are skew scattering\cite{smit} and side-jump scattering\cite{berger} as well as the anomalous velocity operator due to spin-orbit interaction and impurity scattering. A schematic picture of the skew scattering mechanism is shown in Fig. \ref{fig:skew}. After scattering off the impurity potential, there is a spin-dependent probability difference, represented by small angles,  of the electron trajectories. This leads \textit{e.g.} to a slightly larger chance that electrons with spin up moves upwards and electrons with spin down moves downwards after scattering.
The side-jump mechanism is also caused by the combined spin-orbit and impurity scattering, see Fig. \ref{fig:sidejump}. After scattering off the impurity, a small "side-jump", develops between the trajectories of electrons with spin up and down far away from the scattering center.
Additionally, the spin-orbit interaction does not commute with the electron momentum operator. This leads to an anomalous velocity operator that can contribute to the Hall effects, as well.

Spin-orbit scattering is also important in normal metals. It is well known that it causes  a loss of spin coherence. Hirsch predicted the existence of a novel spin Hall effect\cite{hirsh} analogues to the anomalous Hall effect in ferromagnets and developed a phenomenological theory for the effect.  In the absence of spin-orbit scattering, electrons with spin-up and spin-down scatter equally on non-magnetic impurities. However, as seen above, for nonzero spin-orbit interaction, when a current passes through the sample an imbalance between left-moving and right-moving particles is established, and an accompanying transverse spin accumulation potential builds up in the system.  

Zhang computed this spin potential in the diffusive transport regime, and found that it should be measurable \cite{zhang}.  He considered longitudinal transport in a thin normal metal film, and computed the resulting transverse spin Hall voltage. In this regime, the spin-accumulation is weak and he found that the spin Hall voltage is governed by the anomalous velocity operator. By using the framework developed by Hirsch and Zhang, effects of contact resistances on the spin Hall effect have also been considered\cite{Hu:prb03}.

Spin-orbit scattering is also important in $n$- and $p$-doped semiconductors, where Rashba-type spin-orbit coupling\cite{soRashba} leads to interesting spin Hall effects\cite{sogap,semi_so1,semi_so2,semi_so3,semi_so4,semi_so5,semi_so6,semi_so7,semi_so8,semi_so9,semi_so10,semi_so11,semi_so12,murakami,Hu}. The most interesting case is spin transport in a two-dimensional electron system (2DES). The study of this effect is controversial. The debate\cite{semi_so4,semi_so5,semi_so7,semi_so8,semi_so9,semi_so10,semi_so12,mishchenko} is focused on whether or not the spin Hall conductivity has a universal value $\sigma_{\text{sH}}= e/(8 \pi \hbar)$ and in what regime this result is applicable. Our study is complementary to these studies of spin Hall effects in semiconductors. In our normal metal case, the extrinsic\cite{khae} spin Hall effects arise due to the spin-orbit at impurities. In semiconductors, spin-orbit interaction can be important even in ballistic systems\cite{nonimpuresemi}, in the absence of impurities, for systems with broken spatial inversion symmetry. 

In this work, we derive the spin Hall effects in the presense of spin diffusion from the Eilenberger equation\cite{schwab} in presence of spin-orbit coupling\cite{sinitsyn}, magnetic impurities and non-magnetic impurities. We use the Keldysh Green function technique in the quasiclassical approximation. Our calculations go beyond the assumptions in Ref. \onlinecite{zhang}, which only included effects of the anomalous velocity operator, in that we rigorously derive diffusion equations that also include effects of skew scattering and side-jump scattering. Our results agree with the results by Zhang in the limit he considered, where skew scattering and side-jump scattering can be disregarded. We also consider spin and charge transport in normal metals in another transport regime, phenomenologically treated by Hirsch, when normal metals are biased by ferromagnets, in which spins accumulate in the normal metal even in the absence of the spin-orbit interaction. We demonstrate that in this regime, skew scattering and side-jump scattering cannot be disregarded, and compute the spin Hall and spin-orbit induced charge Hall effects. In our analysis, we also consider transport through ferromagnet-normal metal interfaces beyond the assumptions in Ref. \onlinecite{Hu:prb03} by using the boundary conditions obtained within magneto-electronic circuit theory, which is necessary in dealing with non-collinear spin and magnetization directions. Finally, our formalism also explicitly incorporates the effects of scattering off magnetic impurities, which can reduce the spin Hall effects.

Our paper is organized in the following way. In the next section, Section \ref{normet}, we outline our starting point microscopic Hamiltonian, and explain the diffusion equations for spin and charge flow that we have obtained rigorously by using the non-equilibrium Keldysh Green function approach in the quasi-classical approximation. We compute from the diffusion equations the spin Hall effects in some relevant experimental geometries in Section \ref{exper}.  Our derivation of the charge and spin diffusion equation is an important part of our work, and its details are given in Section \ref{micro}. Finally, we give our conclusions in Section \ref{con}.

\section{Model and transport equations}\label{normet}

We consider quasi-particles in a normal metal that interact with non-magnetic and magnetic impurities and include the spin-orbit interaction. The Hamiltonian of the system is
\begin{equation}\label{II-10}
H=-\frac{\hbar ^2}{2m}\nabla ^2+V_{\text{imp}}+ \hat{V}_{\text{so}} + \hat{V}_{\text{sm}} \, ,
\end{equation}
where we below will introduce the various terms. Impurity scattering is represented by the short-range potential
\begin{equation}
V_{\text{imp}} = \sum_i\gamma_i\delta(\bm{r}-\bm{r_i}),
\label{Vimp}
\end{equation}
where $\bm{r_i}$ is the coordinate of the $i$ -th impurity center and $\gamma_i$ is the strength of the scattering potential. It is 
assumed that the scatterers are disordered so that $<V_{\text{imp}}(\bm{r})>=0$ and $<V_{\text{imp}}(\bm{r})V_{\text{imp}}(\bm{r'})>=\delta (\bm{r}-\bm{r'})n\gamma ^2$, 
where $n$ is the impurity density and $\gamma ^2$ is the average fluctuation of the scattering strengths. 

The spin-orbit interaction is described by the Hamiltonian,
\begin{equation}\label{II-20}
\hat V_{\text{so}}=\frac{1}{2}\left[\frac{\alpha }{\hbar k_F^2}\left(\bm{\hat \sigma }\times \bm{\nabla }V_{\text{imp}}\right)\bm{p}\ + \text{h.c.} \right],
\end{equation}
where $\alpha$ is the dimensionless spin-orbit coupling constant, $k_F$ is the Fermi wave number and $\text{h.c.}$ denotes the hermitian conjugate.

Magnetic impurites are introduced by
\begin{equation}\label{II-21}
\hat V_{\text{sm}} = V_{\text{sm}}(\bm{r}) \bm{\hat \sigma} \cdot \bm{S} (\bm{r}), 
\end{equation}
where $V_{\text{sm}}(\bm{r})$ is the strength of the coupling of the itinerant electron spin to the spin of the magnetic impurity $\bm{S} (\bm{r})$.

We are interested in the transport properties of diffusive system where the system size is much larger than the mean free path. A 
rigorous method  to obtain the correct diffusion equation is to start from a microscopic description using the non-equilibrium Keldysh 
formalism. We consequently employ the Keldysh approach with two approximations. First, we consider the quasiclassical approximation, 
which is valid on length scales much larger than the Fermi wavelength, $L\gg \lambda _F$. Second, we use the diffusion approximation which
is valid when the system size is much larger than the mean free path, $L\gg l=v_F\tau $.

The full derivation of the diffusion equation is an important part of the present paper, but it is technically complicated and in order to make the paper more easily accessible we delay its derivation to the interested readers in Section \ref{micro}. First, we show and explain the spin and charge diffusion equations as well as the expression for the corresponding currents that we obtain. We introduce charge and spin-
distributions $\mu _c$ and $\bm {\mu _{\text s}}$, so that the charge density and spin-density are
\begin{eqnarray}
& n(\bm{r})=N_o\mu _c(\bm{r}), {}\nonumber\\
& \bm{s}(\bm{r})=N_o\bm {\mu _{\text s}}(\bm{r}), \nonumber &
\end{eqnarray}
where $N_o$ is the density of states. After considerably algebra outlined in Section \ref{micro}, we find that the resulting diffusion equation for the charge distribution functions is simply
\begin{equation}\label{II-30}
\bm{\nabla ^2}\mu _c=0 \, .
\end{equation}
Similarly, the spin-distribution for small spin-orbit interactions, $\alpha \ll 1$, is governed by
\begin{equation}\label{II-50}
D\bm{\nabla ^2\mu _{\text s}}= \left[ \frac{1}{\tau _{\text{so}}} + \frac{1}{\tau _{\text{sm}}} \right] \bm \mu_{\text{s}} \, ,
\end{equation}
where $D=\frac{1}{3}v_F^2\tau$ is the diffusion coefficient in terms of the Fermi velocity $v_F$ and the elastic scattering time $\tau$, $\tau _{\text{so}}$ is the spin-flip relaxation time due to the spin-orbit interaction, 
\begin{equation}
\quad \frac{1}{\tau _{so}}=\frac{8\alpha ^2}{9\tau }
\label{tau_so}
\end{equation}
and $\tau_{\text{sm}}$ is the spin-flip relaxation time due to magnetic impurities, 
\begin{equation}
\frac{1}{\tau_{\text{sm}}} = \frac{8 \pi n_{\text{sm}} N_0 S(S+1) v_{\text{sm}}^2}{3} \, .
\end{equation}
Here we have expressed the strength of the magnetic impurity potential from Eq. (\ref{II-21}) in the momentum representation $v_{\text sm}$, 
$n_{\text sm}$ is the concentration of the magnetic impurities, and $S$ is the spin of the impurity. Thus, both the diffusion equations for charge and spin have the familiar forms used exensively in the literature for spin and charge transport.

Skew scattering, side-jump scattering and effects of the anomalous velocity operator are all contained in the expressions for the current. We find that the total $2 \times 2$ current in spin-space can be expressed as
\begin{equation}\label{II-60}
\hat{\bm{j}}=\hat{\bm{j}}_{\text{o}}+\hat{\bm{j}}_{\text{av}}+\hat{\bm{j}}_{\text{ss}}+\hat{\bm{j}}_{\text{sj}},
\end{equation}
where $\hat{\bm{j}}_{\text{o}}$ is the ordinary current without spin-orbit interaction, $\hat{\bm{j}}_{\text{av}}$ is the current due to the anomalous velocity operator,  $\hat{\bm{j}}_{\text{ss}}$ is the current due to skew scattering and $\hat{\bm{j}}_{\text{sj}}$ is the current due to side-jump scattering. The full derivation of these currents is given in Section \ref{micro}. The charge and spin currents can be obtained by the trace of Eq. (\ref{II-60}) with the unit matrix and the Pauli matrices, respectively. In the limit of weak spin-orbit interaction (to lowest order in $\alpha $), which is relevant for most normal metals, we compute the following contributions to the current:
\begin{eqnarray}
e\hat{\bm{j}}_{\text{o}} & = & -\sigma \frac{1}{2}\left[ \hat 1\bm{\nabla }\mu _c+\bm{\nabla }(\bm{\mu }_{\text {s}}\hat{\bm{\sigma }}) \right] \label{II-80} \, , \\
e\hat {\bm{j}}_{\text{av}} & = & \sigma \frac{\alpha \hbar }{6mD}\left[\bm{\nabla }\times \bm{\mu }_{\text{s}}+\hat{\bm{\sigma }}
\times \bm{\nabla }\mu _c\right] \label{II-70} \, , \\
e\hat{\bm{j}}_{\text{sj}} & = & \sigma \frac{\alpha }{3}(\hat{\bm{\sigma }}\bm{\nabla })\bm{\mu }_{\text{s}} \label{II-90} \, , \\
e\hat{\bm{j}}_{\text{ss}} & = & -\sigma \alpha \hat{\bm{\sigma }}(\bm{\nabla \mu }_{\text{s}}) \label{II-100} \, ,
\end{eqnarray} 
where $\sigma =e^2 N_0 D$ is the conductivity. 
The anomalous current, $\hat{\bm{j}}_{\text{av}}$, contributes both to the spin and charge current. Contributions from skew scattering and side-jump scattering, $\hat{\bm{j}}_{\text{ss}}$ and $\hat{\bm{j}}_{\text{sj}}$, only affect the spin-current. The current contribution $\hat{\bm{j}}_{\text{sj}}$ depends on the divergence of the spin accumulation in the direction of the current for any spin, which is the side-jump mechanism. The current contribution $\hat{\bm{j}}_{\text{ss}}$ arises from skew scattering.
If we compare our results with Zhang \cite{zhang}, we see that in addition to the contributions from the anomalous velocity operator we include terms representing skew scattering and side-jump scattering. We demonstrate below that these additional contributes could correctly be disregarded in the geometry in Ref. \onlinecite{zhang}, but that they the dominate spin Hall effects in other systems. We also derive expressions for the total current in case of arbitary $\alpha $ (see Section \ref{micro}). 
\section{Experimental implication.} \label{exper}
Let us now employ our theory to calculate the magnitude of the spin Hall effect in experimental relevant geometries. For simplify, we will consider the cases with small spin-orbit interaction ($\alpha \ll 1$), where the small $\alpha$ expressions for the current, (\ref{II-80}), (\ref{II-70}), (\ref{II-90}) and (\ref{II-100}), are valid. 
\subsection{Thin metallic film.}
\begin{figure}[ht]
\includegraphics[angle=0,width=7cm]{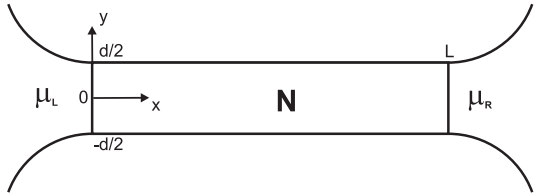}
\caption{The thin metallic film with a contact to reservoir.}
\label{fig:normal}
\end{figure}

We consider first a pure normal metal as considered in Ref. \onlinecite{zhang} and shown in Fig. \ref{fig:normal}. A thin film normal metal of length $L$ and width $d$ is attached with perfect contacts with zero resistance to a left reservoir with local chemical potential $\mu_{\text{L}}$ and a right reservoir with local chemical potential $\mu_{\text{R}}$. In a pure normal metal system, there is no spin-accumulation in the limit $\alpha \rightarrow 0$. That means that the spin-accumulation is small, being induced by the spin-orbit interaction. From (\ref{II-70}), (\ref{II-90}) and (\ref{II-100}) we thus see that contributions from the skew scattering and side-jump scattering to the current are of a higher order in the spin-orbit scattering than the anomalous current and can be disregarded. In the case of pure normal metals, in the weak spin-orbit interaction limit, the current can thus simply be expressed in terms of
\begin{equation}
\hat{\bm{j}} \approx \hat{\bm{j}}_0 +\hat{\bm{j}}_{\text{av}} \, ,
\label{jtotapprox}
\end{equation}
where the anomalous current simplifies to
\begin{equation}
e\hat{\bm{j}}_{\text{av}} \approx \sigma \frac{\alpha \hbar }{6mD}\hat {\bm{\sigma}} \times \bm{\nabla} \mu_c \, .
\label{jacapprox}
\end{equation}
The effect of the spin-orbit interaction is consequently to induce a transverse spin Hall potential. The magnitude of the spin Hall effect depends on the system size and geometry. If the system size is smaller than the spin-diffusion length, a spin accumulation cannot build up within the system, and consequently the spin Hall effect vanishes. 
Therefore, we consider the situation when the system size is much larger than the spin-diffusion length, $L\gg l_{\text sf}=\sqrt{D\tau_{\text s}}$, where the total spin-flip relaxation time has contributions both due to spin-orbit scattering and magnetic impurity scattering, $1/{\tau }_{\text s}=1/{\tau }_{\text so}+1/{\tau }_{\text sm}$. The solution of the diffusion equation is similar to the treatment in Ref. \onlinecite{zhang}. At distances larger than the spin-diffusion length from the reservoirs, the spin potential only depend on the transverse $y$-coordinate. In this regime, the general solution of the diffusion equation (\ref{II-50}) has the form
\begin{equation}\label{III-10}
\bm{\mu }_{\text{s}}=\bm{c}_3e^{y/l_{\text{sf}}}+\bm{c}_4e^{-y/l_{\text{sf}}} \, ,
\end{equation} 
where $\bm{c}_3$ and $\bm{c}_4$ are constants to be determined. %
We can determine the constants $\bm{c}_3$ and $\bm{c}_4$ from the boundary condition that there is no particle or spin flow across the transverse boundaries, \textit{e.g.} the $2\times 2$ current in spin-space must satisfy:
\begin{equation}\label{III-40}
\hat j_y(x,y=\pm d/2)=0 \, .
\end{equation}
Using the simplified equations (\ref{jtotapprox}) and (\ref{jacapprox}) determining the  current, we find $\bm{c}_3$ and $\bm{c}_4$ and the spatially dependent transverse spin potential:
\begin{equation}\label{III-50}
\mu _{s,z}^{(A)}(y)= \sqrt{\frac{3}{8}}\frac{\mu _R-\mu _L}{k_FL}  \sqrt{\frac{\tau_{\text{sm}}}{\tau_{\text{sm}} + \tau_{\text{so}} }}  \frac{{\text{sinh}}(y/l_{\text{sf}})}{{\text{cosh}}(d/2l_{\text{sf}})} \, .
\end{equation} 
This expression exactly agrees with the result of Zhang\cite{zhang} $\mu _{s,z}\equiv \mu ^{\uparrow }-\mu ^{\downarrow }= l_{\text{sf}}E_x \, C_h/C \, {\text{sinh}}(y/l_{\text{sf}})/{\text{cosh}}(d/2l_{\text{sf}})$
when we insert the Drude conductivity $C=e^2\tau (k_F)^3/6\pi ^2m$, the anomalous Hall conductivity $C_h=e^2\alpha _o(k_F)^3/6\pi ^2$ (the dimensionless $\alpha =\alpha _o\hbar k_F^2$) and the electric field in the $x$-direction $E_x=(\mu _R-\mu _L)/L$. In addition, we use the identification between the spin-orbit spin-flip relaxation time and elastic scattering time, Eq. (\ref{tau_so}).

Formula (\ref{III-50}) expresses the transverse spin Hall effect for thin metallic films with small spin-orbit interaction. There are accumulation of spins
directed perpendicular to the film. As we see, the spin-Hall effect vanishes when scattering by magnetic impurities dominates the spin-orbit scattering, $\tau_{\text{sm}} \ll \tau_{\text{so}}$, as expected. When magnetic impurity scattering is weak, and in the limit of a wider film than the spin-diffusion length $d\gg l_{\text{sf}}$, we make the observation that the magnitude of spin Hall effect is "universal", $\mu _{\text{sH}}^{(A)}\equiv \mu _{s,z}^{(A)}(d/2)-\mu _{s,z}^{(A)}(-d/2)$:
\begin{equation}\label{III-60}
\mu_{\text{sH}}^{(A)} \approx \sqrt{\frac{3}{2}}\frac{\mu _R-\mu _L}{k_FL}, 
\end{equation} 
\textit{e.g.} the spin Hall voltage does not depend on the spin-orbit interaction constant $\alpha$. By "universal", we mean that the spin Hall potential does not depend on the strength of the spin-orbit scattering potential. This implies that, as long as the scattering off magnetic impurities is weak, light metals (\textit{e.g} Cu, Al) will give a similar spin Hall voltage as heavy metals (\textit{e.g} Pt). Note that the spin Hall potential depend on the Fermi wave vector of the metal, $k_F$, and thus weakly depends on the type of normal metal through this dependence. The reason for the "universality" is that although the spin-Hall potential is proportional to the spin-orbit scattering it only builds up within the spin-diffusion length which is inversely proportional to the spin-orbit scattering strength.

This section also illustrates the differences between the extrinsic spin-orbit scattering off impurities in normal metals, as treated here, and spin-orbit scattering induced by the Rashba term and impurities in the two-dimensional electron gas formed in semiconductor heterostructures. In the two-dimensional electron gas, there is currently a controversy whether the spin Hall conductivity can reach a universal value in dirty systems, $\sigma_{\text{sH}}= e/(8 \pi \hbar)$, independent of the spin-orbit scattering strength. In normal metals, we see from (\ref{jacapprox}) that the spin Hall conductivity is not universal, but the spin Hall voltage (\ref{III-60}) can be "universal" when scattering off magnetic impurities is weak.

\subsection{Thin metallic film in tunneling contacts to ferromagnet and normal metal.}

\begin{figure}[ht]
\includegraphics[angle=0,width=7cm]{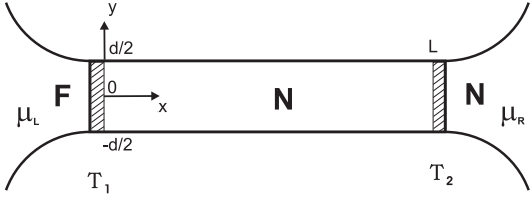}
\caption{The thin metallic film with tunneling contacts $T_1$ and $T_2$ to ferromagnet and normal metal.}
\label{fig:fig4}
\end{figure}

We will now consider another transport regime, in which there is a finite spin-accumulation present in the normal metal even in the absence of spin-orbit scattering.  The spin-accumulation can be achieved by sandwiching the normal metal between two ferromagnets when the system is driven out of equilibrium, see Fig. \ref{fig:fig4}.
There are two extreme regimes depending on the ratio between the system length and the spin-diffusion length. The first regime, when the length is large, ($L\gg l_{\text{sf}}$), does not provide novel physics since it is similar to the previous example. In this case any spins injected from the ferromagnets will be lost, and the system resembles a pure normal metal where a spin Hall voltage can build up transverse to the current direction. There is no charge Hall effect produced by the spin-orbit interaction in the middle of the system in this case. Therefore, we consider the second,  more interesting regime, when the normal metal is short, ($L\ll l_{\text{sf}}$). In this case both spin and charge Hall potentials can build up transverse to the current direction governed by the spin-orbit scattering. 

In order to further simplify the computation of the diffusion equations (\ref{II-30}), (\ref{II-50}) we assume that the normal metal is narrow, \textit{e.g.} $d\ll l_{\text{sf}}$ as well. The diffusion equations then simplify to
$\bm{\nabla ^2}\mu _c=0$ and $\bm{\nabla ^2\mu }_{\text{s}}=0$.

In addition, we need the boundary condition of the spin and charge flow through the tunnel contacts from the ferromagnetic reservoirs into the normal metal wire. These boundary conditions follow from magnetoelectronic circuit theory, so that the interface transport can be described by spin-dependent conductances $G^{\uparrow}$ and $G^{\downarrow}$ for spin aligned and antialigned to the magnetization and a mixing conductance of reflection $G^{\uparrow \downarrow}$ for spins in the normal metal that are non-collinear to the magnetization direction.\cite{arne} For most systems, ${\text{Re}}G^{\uparrow \downarrow }\gg {\text{Im}}G^{\uparrow \downarrow }$, and this will be assumed in the following, which simplifies the expressions for charge and spin currents\cite{arne}. The charge current along the transport, $x$, direction through the tunneling contact can be written as
\begin{equation}\label{III-80}
eI_c=(G^{\uparrow }+G^{\downarrow })\left[\mu _c^F-\mu _c(0,y)\right]-(G^{\uparrow }-G^{\downarrow })\bm{m}\cdot \bm{\mu }_{\text{s}}(0,y)
\end{equation}
in terms of the local quasi-equilibrium chemical potential in the ferromagnet $\mu _c^F$ and the spin and charge chemical potentials in the normal metal close to the ferromagnetic interface $\bm{\mu }_{\text{s}}(x=0,y)$ and $\mu _{\text{c}}(x=0,y)$. Here $\bm{m}$ is the magnetization of the ferromagnet ($|\bm{m}|=1$). For a tunnel contact $2{\text{Re}}G^{\uparrow \downarrow }=G^{\uparrow }+G^{\downarrow }$.\cite{arne} The spin-current along the transport, $x$, direction is then
\begin{eqnarray}\label{III-100}
& e\bm{I}_{\text{s}}=\bm{m}[(G^{\uparrow }-G^{\downarrow })(\mu _c^F-\mu _c(0,y))+(G^{\uparrow }+G^{\downarrow })\mu _s^F]- {}\nonumber\\
& 2{\text{Re}}G^{\uparrow \downarrow }\bm{\mu }_{\text{s}}(0,y). &
\end{eqnarray}

We assume that the spin-orbit interaction is weak, and expand the charge and spin accumulations to first order in $\alpha $:
\begin{eqnarray}
{\mu }_c(x,y) & = & {\mu }_c^{(0)}(x,y)+\alpha \delta {\mu }_c(x,y) {}\label{III-61} \, , \\
\bm{\mu }_{\text{s}}(x,y) & = & \bm{\mu }_{\text{s}}^{(0)}(x,y)+\alpha \delta \bm{\mu }_{\text{s}}(x,y) {}\label{III-70} ,
\end{eqnarray}
where the zeroth order terms are the charge and spin accumulations in the absence of spin-orbit interaction and the corrections are caused by the spin-orbit interaction. 
First, we compute the zeroth order terms that correspond to $\alpha =0$. The boundary condition $\hat j_y^0(x,y=\pm d/2)=0$ of no current out of the transverse edges dictates that the general solutions $\mu _c^{(0)}$ and $\bm{\mu }_{\text{s}}^{(0)}$ only vary in the transport direction: 
$\mu _c^{(0)}=c_0+c_1\frac{x}{L}$, $\bm{\mu }_{\text{s}}^{(0)}=\bm{c}_{\text{o}}+\bm{c}_1\frac{x}{L}$. 

Now we can equate the current through the tunnel contact with the current in normal metal at $x=0$. This is a boundary condition on the ferromagnet $T_1$ 
(see Fig.4). For the charge current 
\begin{equation}\label{III-110}
-G_oc_1=(G^{\uparrow }+G^{\downarrow })(\mu _L-c_0)-(G^{\uparrow }-G^{\downarrow })\bm{m}\cdot \bm{c}_0,
\end{equation}
where the conductance of the normal metal is $G_o=\sigma wd/L$, $w$ is the width of the film. For spin-current
\begin{equation}\label{III-120}
-G_o\bm{c}_1=-(G^{\uparrow }+G^{\downarrow })\bm{c}_0+\bm{m}[(G^{\uparrow }-G^{\downarrow })(\mu _L-c_0).
\end{equation}
At the normal metal-normal metal tunnel contact $T_2$ (see Fig.4) we use the same expressions for currents (\ref{III-80}),(\ref{III-100}) with $G=G^{\uparrow }=G^{\downarrow }$, and the boundary conditions are
\begin{eqnarray}
& G(c_0+c_1-\mu _R)=-G_oc_1 {}\label{III-130},\\
& G(\bm{c}_0+\bm{c}_1)=-G_o\bm{c}_1. \label{III-140} &
\end{eqnarray}
After straightforward calculations we find 
\begin{equation}\label{III-150}
\bm{\mu }_{\text{s}}^{(0)}(x)=\left[1-\frac{G}{G+G_o}\frac{x}{L}\right]\frac{G^{\uparrow }_{\text{tot}}-G^{\downarrow }_{\text{tot}}}{G^{\uparrow }_{\text{tot}}+G^{\downarrow }_{\text{tot}}}(\mu _L-\mu _R)\bm{m},
\end{equation}
\begin{equation}\label{III-160}
\mu _c^{(0)}(x)=c_0-\frac{4}{G_o}\left(\frac{1}{G^{\uparrow }_{\text{tot}}}+\frac{1}{G^{\uparrow }_{\text{tot}}}\right)(\mu _L-\mu _R)\frac{x}{L},
\end{equation}
where the totale spin-dependent conductances of the system are given by the resistances in series $1/G^{\uparrow (\downarrow )}_{\text{tot}}=1/G^{\uparrow (\downarrow )}+1/G+1/G_o$. $c_0$ can be expressed similarly, but is not shown here since it does not govern the spin-orbit induced correction to the currents which will be considered next.

Next we introduce spin-orbit interaction which will produce current correction $\delta \hat {\bm{j}}(x,y)$ so that the full current is expressed as  $\hat {\bm{j}}=\hat {\bm{j}}^0+\delta \hat {\bm{j}}$. The boundary condition $\delta \hat j_y(y=\pm d/2)=0$ gives
\begin{eqnarray}
\frac{\partial }{\partial y}\delta \mu _c\mid _{y=\pm d/2} & = & -\frac{\hbar }{3mD}\frac{\partial }{\partial x}\delta \bm{{\mu }}_{\text{s},z}^{(0)}\label{III-161}\\
\frac{\partial }{\partial y}\delta \mu _{{\text s},x}\mid _{y=\pm d/2} & = & \frac{2}{3}\frac{\partial }{\partial x}\delta \bm{{\mu }}_{\text{s},y}^{(0)}\label{III-162}\\
\frac{\partial }{\partial y}\delta \mu _{{\text s},y}\mid _{y=\pm d/2} & = & -2\frac{\partial }{\partial x}\delta \bm{{\mu }}_{\text{s},x}^{(0)}\label{III-163}\\
\frac{\partial }{\partial y}\delta \mu _{{\text s},z}\mid _{y=\pm d/2} & = & \frac{\hbar }{3mD}\frac{\partial }{\partial x}\delta \bm{{\mu }}_c^{(0)}\label{III-164}
\end{eqnarray}
where we have introduced the total polarization of the conductance of the system $p_{\text{tot}}\equiv (G^{\uparrow }_{\text{tot}}-G^{\downarrow }_{\text{tot}})/(G^{\uparrow }_{\text{tot}}+G^{\downarrow }_{\text{tot}})$. The derivative of $x$-component of the correction to the spin potential is governed by side-jump current, $y$-component by skew scattering current and $z$-component by anomalous current. In the limit $d\ll L$ we can expand the corrections to the chemical potentials in the small parameter $(d/L)$, so that the solutions can be represented as 
\begin{eqnarray}
\delta {\mu }_c & = & \delta {\mu }_c^{(0)}+\frac{d}{L}\delta {\mu }_c^{(1)} {}\label{III-280}\\
\delta {\bm{\mu }}_{\text{s}} & = & \delta \bm{{\mu }}_{\text{s}}^{(0)}+\frac{d}{L}\delta \bm{{\mu }}_{\text{s}}^{(1)}\label{III-290}
\end{eqnarray}
Due to simplified diffusion equations we write 
\begin{eqnarray}
\delta {\mu }_c^{(i)} & = & c_0^{(i)}+c_1^{(i)}x+c_2^{(i)}y, {}\label{III-290}\\
\delta \bm{{\mu }}_{\text{s}}^{(i)} & = & \bm{c}_0^{(i)}+\bm{c}_1^{(i)}x+\bm{c}_2^{(i)}y. \label{III-300}
\end{eqnarray}  
where $i=0,1$. We are interesting for coefficients $c_2^{(i)}$ and $\bm{c}_2^{(i)}$ because $\alpha c_j^{(i)}\ll c_j/L$, $\alpha \bm{c}_j^{(i)}\ll \bm{c}_j$; $j=0,1$. The boundary condition on the contact $T_1$ gives 
\begin{eqnarray}
\frac{\partial }{\partial x}\delta \mu _c\mid _{x=0} & = & \frac{G^{\uparrow }+G^{\downarrow }}{G_o}\delta \mu _c(0,y)+\frac{G^{\uparrow }-G^{\downarrow }}{G_o}\bm{m}\delta \bm{{\mu }}_{\text{s}}(0,y), {}\nonumber\\
\frac{\partial }{\partial x}\delta \bm{{\mu }}_{\text{s}}\mid _{x=0} & = & \frac{G^{\uparrow }-G^{\downarrow }}{G_o}\delta \mu _c(0,y)\bm m+\frac{G^{\uparrow }+G^{\downarrow }}{G_o}\delta \bm{{\mu }}_{\text{s}}(0,y)- {}\nonumber\\
& & \frac{4}{3}\frac{\partial }{\partial x}\delta \mu _{{\text s},x}^{(0)}\bm{e}_x, {}\nonumber
\end{eqnarray}
and similarly for the contact $T_2$. If we consider these conditions for $\delta {\mu }_c\cong \delta {\mu }_c^{(0)}$, $\delta \bm{{\mu }}_{\text{s}}\cong \delta \bm{{\mu }}_{\text{s}}^{(0)}$ we derive $c_2^{(0)}=\bm{c}_2^{(0)}=0$. Conditions (\ref{III-161})-(\ref{III-164}) give
\begin{eqnarray}
c_2^{(1)} & = & \frac{\hbar m_z}{3mD}\frac{G\,p_{\text{tot}}}{G+G_o}\frac{\mu _L-\mu _R}{d} {}\nonumber\\
c_{2,x}^{(1)} & = & -\frac{2}{3}m_y\frac{G\,p_{\text{tot}}}{G+G_o}\frac{\mu _L-\mu _R}{d} {}\nonumber\\
c_{2,y}^{(1)} & = & 2m_x\frac{G\,p_{\text{tot}}}{G+G_o}\frac{\mu _L-\mu _R}{d} {}\nonumber\\
c_{2,z}^{(1)} & = & -\frac{\hbar }{3mD}\frac{4}{G_o}\frac{G^{\uparrow }_{\text{tot}}G^{\downarrow }_{\text{tot}}}{G^{\uparrow }_{\text{tot}}+G^{\downarrow }_{\text{tot}}}\frac{\mu _L-\mu _R}{d}, {}\nonumber
\end{eqnarray}
Finally, we have for the spin-orbit induced charge Hall effect
\begin{equation}
\delta \mu _c^{(B)}(x,y)  \approx   \alpha m_z\frac{G\,p_{\text{tot}}}{G+G_o}\frac{\mu _L-\mu _R}{k_FL}\frac{y}{l}, \label{III-310}
\end{equation}
and for the spin Hall effect
\begin{eqnarray}
\delta \bm{{\mu }}_{\text{s},x}^{(B)}(x,y) & \approx & -\frac{2}{3}\alpha m_y\frac{\mu _L-\mu _R}{L}\frac{G\,p_{\text{tot}}}{G+G_o}y, \label{III-321}\\
\delta \bm{{\mu }}_{\text{s},y}^{(B)}(x,y) & \approx & 2\alpha m_x\frac{\mu _L-\mu _R}{L}\frac{G\,p_{\text{tot}}}{G+G_o}y, \label{III-322}\\
\delta \bm{{\mu }}_{\text{s},z}^{(B)}(x,y) & \approx & -\frac{4\alpha }{k_FlG_o}\frac{\mu _L-\mu _R}{L}\frac{G^{\uparrow }_{\text{tot}}G^{\downarrow }_{\text{tot}}}{G^{\uparrow }_{\text{tot}}+G^{\downarrow }_{\text{tot}}}y, \label{III-323}
\end{eqnarray}
where $l$ is the electron mean free path. As we see from the equation (\ref{III-310}) the spin-orbit induced charge Hall effect is non-zero only for nonzero magnetization of the ferromagnet in the $z$-direction 
(direction which is perpendicular to the film). At the same time the spin-Hall effect of the spins along $z$, (\ref{III-323}), is independent of the magnetization direction. So, assuming $\bm{m}=\{0,0,1\}$ the magnitude of Hall effect is
\begin{equation}\label{III-325}
\mu _H^{(B)}\approx -\alpha \frac{d}{l}\frac{G\,p_{\text{tot}}}{G+G_o}\frac{\mu _R-\mu _L}{k_FL},
\end{equation}
and the spin Hall effect
\begin{equation}\label{III-330}
\mu _{sH}^{(B)}\approx \alpha \frac{d}{l}\frac{4}{G_o}\frac{G^{\uparrow }_{\text{tot}}G^{\downarrow }_{\text{tot}}}{G^{\uparrow }_{\text{tot}}+G^{\downarrow }_{\text{tot}}}\frac{\mu _R-\mu _L}{k_FL} \, .
\end{equation}
The spin-orbit induced charge Hall effect vanishes when the polarization goes to zero, as should be expected.
In comparison, we give the magnitude of spin Hall effect in pure normal metal regime in the limit of $d\ll l_{\text{sf}}$. As directly follows from eq. (\ref{III-50}) 
\begin{equation}\label{III-335}
\mu _{sH}^{(A)}(d\ll l_{\text{sf}})=\alpha \frac{d}{l}\frac{\mu _R-\mu _L}{k_FL} \, .
\end{equation} 
To evaluate the expressions (\ref{III-325}), (\ref{III-330}) and compare them to (\ref{III-335}) we assume the conductance at the tunnel barrier $T_2$ equals to the sum of spin-dependent conductances at the barrier $T_1$, so $G=G^{\uparrow }+G^{\downarrow }$ and take a typical value of the polarization of the conductance of the system $p\equiv (G^{\uparrow }-G^{\downarrow })/(G^{\uparrow }+G^{\downarrow })=1/2$. 
The limit $G\gg G_o$ is not interesting because in this case our system will be similar to a pure normal metal attached to reservoirs, which was considered in the previous section. Another limit  $G\ll G_o$ is also less interesting due to the small induced voltage across the normal metal and consequently vanishing Hall effects. So, we consider the most interesting case $G=G_o$. In this case
\begin{eqnarray}
\frac{\mu _{sH}^{(B)}}{\mu _{sH}^{(A)}(d\ll l_{\text{sf}})} & = & \frac{3}{7}\label{III-341} \, , {}\\
\frac{\mu _{H}^{(B)}}{\mu _{sH}^{(A)}(d\ll l_{\text{sf}})} & = & \frac{1}{7}\label{III-342} \, .
\end{eqnarray}
We see from these expressions that both the charge and the spin Hall effects are comparable to the spin Hall effect in the pure normal metal regime attached to leads with perfect normal metal contacts in the regime $L\gg l_{\text{sf}}$. Although the magnitudes of charge and spin Hall effects have a similar structure in both regimes, they have a different origin in principle. The effects in this section are due to skew scattering, side-jump scattering and anomaloys velocity, while the effects in the previous section are due to the anomalous velocity operator only. We treated in this section the regime $d \ll L$. We expect that the spin Hall and the charge Hall effects increase their magnitudes $\mu _{H}^{(B)}$ and $\mu _{sH}^{(B)}$ with increasing of $d$ until saturation when $d \sim l_{\text{sf}}$ similary to the case when the spin Hall effect in pure normal metal attains the "universal value" (\ref{III-60}).

\section{Microscopic derivation of diffusion equation} \label{micro}
We will in this section derive the diffusion equations rigorously from the microscopoic Hamiltonian with the Keldysh Green's function technique in the quasi-classical limit. The nonmagnetic impurity potential (\ref{Vimp}) in terms of incident $\bm{k}$ and scattered $\bm{k'}$ wave vectors is $V_{\text{imp}}(\bm{k},\bm{k'})=\sum_i\gamma_i \exp{-i\left(\bm{k}-\bm{k'}\right)\bm{r_i}}$. Consequently, the spin-orbit interaction in this representation is $\hat V_{\text{so}}(\bm{k},\bm{k'})=-i(\alpha/k_F^2)\bm{\hat \sigma} \cdot \left(\bm{k}\times \bm{k'}\right)V_{\text{imp}}(\bm{k}-\bm{k'})$. Thus, in normal metals with dilute impurities the electrons interact with the potential $V(\bm{k},\bm{k'})\equiv V_{\text{imp}}+V_{\text{so}}$:
\begin{equation}\label{A120}
\hat{V}(\bm{k},\bm{k'})=\hat MV_{\text{imp}}(\bm{k},\bm{k'}),
\end{equation}
where we have introduced the $2\times 2$ matrix in spin-space 
\begin{equation}\label{A110}
\hat M\equiv \hat 1-i\frac{\alpha }{k_F^2}\bm{\hat \sigma}\left(\bm{k}\times \bm{k'}\right) \, .
\end{equation}
In addition, the electrons interact with magnetic impurities (\ref{II-21}) to be discussed below. Our transport theory is based on the Keldysh formalism\cite{schwab,rammer}. In this formalism the Green's function has the form
\begin{equation}\label{A40}
\check G=\left(
\begin{array}{cc}
\hat G^R & \hat G^K \\ 0 & \hat G^A
\end{array}
\right),
\end{equation}
where the retarded, advanced and Keldysh Green's functions are
\begin{eqnarray}
\hat G^R & = & -i\theta (t_1-t_2)\langle[\Psi (x_1),\Psi ^+(x_2)]_+  \rangle  \, , \label{A51} \\
\hat G^A & = & +i\theta (t_2-t_1)\langle[\Psi (x_1),\Psi ^+(x_2)]_+  \rangle  \, , \label{A52} \\
\hat G^K & = & -i\langle[\Psi (x_1),\Psi ^+(x_2)]_- \rangle \, . \label{A53} 
\end{eqnarray}
Here $\Psi $ is the fermion annihilation operator, $\Psi ^+$ is the fermion creation operator, both in the Heisenberg picture, and the coordinate $x_i$ denotes both spatial position and time, $x_i=({\bf x_i}, t_i)$. Note that the fermion annihilation and creation operators are 2-component vectors in spin-space.
The self-energy has the same triangular matrix structure as the Green's function,
\begin{equation}\label{A60}
\check \Sigma =\left(
\begin{array}{cc}
\hat \Sigma ^R & \hat \Sigma ^K \\ 0 & \hat \Sigma ^A
\end{array}
\right) \, .
\end{equation}
We denote $4\times 4$ matrices in Keldysh space by the symbol "check" ($\check{}$) and $2 \times 2$ matrices in spin space by the symbel "hat" ($\hat{}$). Next we define the center-of-mass and relative variables ${\bf x}=\frac{1}{2}({\bf x}_1+{\bf x}_2), \quad {\bf r}={\bf x}_1-{\bf x}_2$, 
and Fourier transform with respect to the relative coordinate ${\bf r}$ in order to obtain the Green's function in the mixed representation
\begin{equation}
\check G({\bf p},{\bf x})=\int d{\bf r}e^{-i{\bf p\cdot r}}\check G({\bf x}+{\bf r}/2,{\bf x}-{\bf r}/2) \, .
\end{equation}
We will also use the $\xi $-integrated (quasiclassical) Green's function $\check g({\bm{n}},\bm{r})=(i / \pi) \int d\xi \check G(\bm{p},\bm{r})$, 
where $\xi =\bm{p}^2/2m-\mu $, and $\bm{n}$ is a unit vector along the momentum at the Fermi surface (${\bm{n}}=\bm{k_F}/|\bm{k_F}|$).
Let us first consider the current produced by the normal velocity operator of the electrons,  
$\hat{\bm{j}}_{\text{{N}}}$. This current is expressed as $\hat{\bm{j}}_{\text{N}}=(e/m) \Re \langle\Psi ^+\bm p\Psi \rangle $, 
where $\bm p$ is the momentum operator. Introducing the Green's functions, the current due to the normal velocity operator is 
\begin{equation}\label{A90}
\hat{\bm{j}}_{\text{N}}(x_1) = -\frac{e}{2m}\langle\lim_{x_1\rightarrow x_2}(\bm{\nabla }_1-\bm{\nabla }_2)\hat G^K(x_1,x_2)\rangle \, .
\end{equation}
Inserting the Fourier representation of the Green's function in the quasiclassical approximation gives for the ordinary current in the mixed representation\cite{schwab}
\begin{equation}\label{A100}
\hat{\bm{j}}_{\text{N}}=\frac{eN_o}{2}\int d\varepsilon \int \frac{d\bm{n}}{4\pi }v_F\bm{n}\, \left< \hat{g}^K(\bm{n},\bm{r})  \right>  \, ,
\end{equation}
where $\left<...\right>$ denotes averaging over impurities and $N_o$ is the density of states at the Fermi level. 

The total current also has contributions caused by the anomalous velocity operator. This contribution to the current can be expressed \cite{kopnin,schwab} as
\begin{equation}\label{A160}
\hat {\bm{j}}_{\text{av}}=\frac{eN_o}{2}\int d\varepsilon \int \frac{d\bm{n}}{4\pi }\frac{d\bm{n'}}{4\pi }\left<
\hat v_{\text{so}}(\bm{n},\bm{n'})\check g(\bm{n'},\bm{r})\right>^K \, .
\end{equation}
The anomalous current caused by the spin-orbit interaction is 
\begin{equation}\label{A130}
\hat v_{\text{so}}(\bm{r})\equiv \frac{dV_{\text{so}}(\bm{r})}{d\bm{p}}=\frac{\alpha }{\hbar k_F^2}\bm{\hat \sigma }\times \bm{\nabla} V_{\text{imp}}(\bm{r}),
\end{equation}
In Fourier space,
\begin{equation}\label{A140}
\hat v_{\text{so}}(\bm{k},\bm{k'})=\frac{\alpha }{\hbar k_F^2}\bm{\hat \sigma }\times \left<k\mid \bm{\nabla} V_{\text{imp}}\mid k'\right>=\hat NV_{\text{imp}}(\bm{k},\bm{k'}).
\end{equation}
where we have introduced the $2\times 2$ matrix in spin-space 
\begin{equation}\label{A141}
\hat N\equiv i\frac{\alpha }{\hbar k_F}\bm{\hat \sigma}\left(\bm{k}-\bm{k'}\right) \, .
\end{equation}

The challenge is now to find the average, $\left<\hat v_{\text{so}}(\bm{n},\bm{n'})\check g\right>$. Note that this is different than the average Green's function appearing in the contribution from the normal velocity operator (\ref{A100}). In the anomalous current (\ref{A130}), a product of the spin-orbit scattering potential \textit{and} the Green's function, both of which depends on the impurity configuration has to be evaluated, and in general one should expect that $\left<\hat v_{\text{so}}\check g\right> \ne \left<\hat v_{\text{so}} \right> \left< \check g\right>$. Using the Dyson equation $\left<\check g\right>=\check g_o+\check g_o\hat \Sigma \left<\check g\right> $ and $\left<\hat v_{\text{so}}\check g\right> = \hat{N} \hat{M}^{-1} \left< \hat{V}\check g\right>$ we find the result
\begin{equation}\label{A200}
\left<v_{\text{so}}\check g\right>=\hat N\hat M^{-1}\Sigma  \left<\check g\right> \, .
\end{equation}
The self-energy part $\hat \Sigma $ can be expressed in the Born approximation as $
\hat \Sigma =\left<\hat{V} \check g \hat{V} \right>$ and has one contribution due to scattering off non-magnetic impurties and two contributions due to the spin-orbit interaction, $\check \Sigma(\bm{n})  =\check \Sigma _{\text{i}}(\bm{n}) +\check \Sigma _{\text{so}}^{(1)}(\bm{n}) +\check \Sigma _{\text{so}}^{(2)} (\bm{n}) +\check \Sigma _{\text{sm}}(\bm{n}) $, where
\begin{eqnarray}
\check \Sigma_{\text{i}}& = & -\frac{i\hbar }{2\tau }\left<\check g(\bm{n'})\right>_{\bm{n'}}\label{A221} \, , \\
\check \Sigma_{\text{so}}^{(2)} & = & -\frac{i\hbar \alpha ^2}{2\tau }\left<\hat{\bm{\sigma }}
(\bm{n}\times \bm{n'})\check g(\bm{n'})\hat{\bm{\sigma }}(\bm{n}\times \bm{n'})\right>_{\bm{n'}}\label{A222} \,  \\
\check \Sigma_{\text{so}}^{(1)} & = & -\frac{\hbar \alpha}{2\tau }\left<\check g(\bm{n'})\hat{\bm{\sigma }}(\bm{n}\times \bm{n'})+ {\text h.c.}
 \right>_{\bm{n'}}\label{A223} \, . 
\end{eqnarray}
Scattering by magnetic impurities (\ref{II-21}) do not cause additional terms in the expressions for the current density, but gives an additional contribution the the electron self-energy
\begin{equation}
\check \Sigma_{\text sm}  =  -\frac{i\hbar }{2\tau _{\text sm}}\frac{1}{3}\sum_{i}\hat \sigma _i\left<\check g(\bm{n'})\right>_{\bm{n'}}\hat \sigma _i ,\label{A224} 
\end{equation}
where the spin-flip relaxation time due to magnetic impurity scattering is
\begin{equation}\label{A225}
\frac{1}{\tau_{\text sm}} = \frac{8 \pi }{3}n_{\text sm} N_o S(S+1) \int {\frac{d\bm{n'}}{4\pi }|V_{\text sm}(\bm{n}-\bm{n'})|^2} \, ,
\end{equation}
$n_{\text sm}$ is the concentration of the magnetic impurities and $S$ is the quantum spin number of the impurity. The self-energy due to scattering off magnetic impurities will leads to simple additional spin-flip relaxation terms in the spin diffusion equation.
In normal metals, the retarded component of Green's function equals $\hat g^R=\hat 1$, and $\hat g^A=-g^R=-\hat 1$. By evaluating the Keldysh component Eq. (\ref{A200}) the anomomalous current is given as 
\begin{eqnarray}\label{A230}
& \hat {\bm{j}}_{\text{av}}=-\frac{ieN_o\hbar }{8\tau }\int d\varepsilon \int \frac{d\bm{n}}{4\pi }\frac{d\bm{n'}}{4\pi }\hat N\hat M^{-1}
[(2+\alpha ^2)\hat g^K(\bm{n'})+ {}\nonumber\\
& \alpha ^2\hat{\bm{\sigma }}(\bm{n}\times \bm{n'})\hat g^K(\bm{n'})\hat{\bm{\sigma }}(\bm{n}\times \bm{n'})
-i\alpha \{\hat g^K(\bm{n'})\hat{\bm{\sigma }}(\bm{n}\times \bm{n'})+ {}\nonumber\\ 
& \hat{\bm{\sigma }}(\bm{n}\times \bm{n'})\hat g^K(\bm{n'})\}] + h.c.&
\end{eqnarray}
We consider transport in the diffuse transport regime. In the diffusive regime, characterized by $v_F\tau \ll L$ ($L$ is the system size) the Green's function is almost isotropic\cite{schwab}, and we can then expand the Green's function in form of isotropic and nonisotropic parts,
\begin{equation}\label{A240}
\int d\varepsilon \hat g^{K}(\bm{r},\bm{n})={\hat \mu }_o(\bm{r})+\hat{\bm{j_1}}(\bm{r})\, \bm{n}.
\end{equation} 
After integrating of the formula (\ref{A230}) we have the result
\begin{widetext} 
\begin{equation}\label{A260}
\hat {\bm{j}}_{\text{av}}= \frac{eN_o}{8\tau k_F}\left\{(\hat{\bm{\sigma }}\times \hat{\bm{j_1}}
-\hat{\bm{j_1}}\times \hat{\bm{\sigma }})f_1+\sum_{i}\bm{e_i}\sum_{jk}\epsilon _{ijk}[\text{Tr}(\hat \sigma _j\hat j_{1,k})f_2+
\text{Tr}(\hat \sigma _k\hat j_{1,j})f_3]\right\}
\end{equation}
where $\text{Tr}(...)$ denotes the sum of diagonal elements of a matrix and the spin-orbit interaction strength dependent functions $f_i(\alpha )$ are
\begin{eqnarray}
f_1 & = & -\alpha +\frac{\alpha ^2}{9} + \alpha (1-\alpha ^2)\int \frac{(n'_z)^2-\alpha \,(n_z)^2(n'_x)^2}{1+\alpha ^2|\bm{n}\times \bm{n'}|^2}
\frac{d\bm{n}}{4\pi }\frac{d\bm{n'}}{4\pi } \label{A271} \\
f_2 & = & \frac{2\alpha }{9} + 2\alpha ^2\int \frac{2\alpha (n_i)^2(n'_j)^2(n'_k)^2-(n_k)^2(n'_j)^2}{1+\alpha ^2|\bm{n}\times \bm{n'}|^2}
\frac{d\bm{n}}{4\pi }\frac{d\bm{n'}}{4\pi } \label{A272} \\
f_3 & = & \frac{2\alpha }{3} + 2\alpha \int \frac{2\alpha (n_i)^2(n'_k)^4+2\alpha ^2(n_k)^2(n'_i)^2(n'_k)^2-(n'_k)^2}{1+\alpha ^2|\bm{n}\times \bm{n'}|^2}
\frac{d\bm{n}}{4\pi }\frac{d\bm{n'}}{4\pi } \label{A273}
\end{eqnarray}
\end{widetext}
In the limit of weak spin-orbit interaction ($\alpha \ll 1$) we thus find the current contribution due to the anomalous veloicty operator
\begin{equation}
\hat{\bm{j}}_{\text{av}} = \frac{\alpha eN_o}{8\tau k_F}\left(\hat{\bm{j_1}}\times \hat{\bm{\sigma }}-
\hat{\bm{\sigma }}\times \hat{\bm{j_1}}\right)
\end{equation}
We have now found the full expression for the current density in terms of the distribution functions and we will now proceed to compute the diffusion equation. The starting point in calculating the Green's function is the equation of motion for the impurity averaged quasiclassical Green's function, the Eilenberger equation. 
\begin{equation}\label{A280}
\hbar \bm{v_F}\bm{\nabla }\langle \check g\rangle +i\left[\check \Sigma ,\langle \check g\rangle \right]_-=0,
\end{equation}
where $\bm{v_F}$ is Fermy velosity. We will in the following omit the impurity average symbol and only implicity write averaging over momentum $\hat{\bm{k}}$ or $\hat{\bm{k'}}$.
We calculate Keldysh components of the various self-energy commutators. For scattering off non-magnetic impurities we obtain the well-known result:
\begin{equation}\label{A290}
i\left[\check \Sigma _{\text i},\check g\right]^K=\frac{\hbar }{\tau }\left(\hat g^K-\left<\hat g^K\right>\right)
\end{equation}
For spin-orbit scattering, which gives rise to side-jump and skew-skattering, using $\left<\hat{\bm{\sigma }}(\bm{n}\times \bm{n'})\hat{\bm{\sigma }}(\bm{n}\times \bm{n'})\right>_{\bm{n'}}=\hat 1$ we find
\begin{equation}\label{A310}
\hat \Sigma _{\text so}^{(2)R}=-\frac{i\hbar \alpha ^2}{2\tau }\hat 1,\quad \hat \Sigma _{\text so}^{(2)A}=\frac{i\hbar \alpha ^2}{2\tau }\hat 1
\end{equation}
and 
\begin{equation}\label{A320}
i\left[\check \Sigma _{\text so}^{(2)},\check g\right]^K=\frac{\hbar \alpha ^2}{\tau }\left[\frac{2}{3}\hat g^K+\left<\hat{\bm{\sigma }}
(\bm{n}\times \bm{n'})\hat g^K(\bm{n'})\hat{\bm{\sigma }}(\bm{n'}\times \bm{n})\right>\right].
\end{equation}
Similarly, using $\left<\hat{\bm{\sigma }}(\bm{n}\times \bm{n'})\right>=0$, we find $\hat \Sigma _{\text so}^{(1)R}=\hat \Sigma _{\text so}^{(1)A}=0$ and
\begin{equation}\label{A340}
i\left[\check \Sigma _{\text so}^{(1)},\check g\right]^K=\frac{i\hbar \alpha}{\tau }\left<\hat g^K(\bm{n'})
\hat{\bm{\sigma }}(\bm{n}\times \bm{n'})-\hat{\bm{\sigma }}(\bm{n}\times \bm{n'})\hat g^K(\bm{n'})\right>
\end{equation}
In the diffusive transport regime, we use the representation 
\begin{equation}\label{A350}
{\hat \mu }_o=\mu _c\hat 1+\bm{\mu _{\text s}}\hat{\bm{\sigma }} \, .
\end{equation}
Employing (\ref{A310}), we find after averaging over momentum, the expression for $\hat \Omega _{so}^{(i)}\equiv i\int d\varepsilon \left[\check \Sigma _{\text so}^{(2)},\check g\right]^K$, where $\varepsilon $ is the energy spectrum of the system:
\begin{eqnarray}
\hat \Omega _{\text so}^{(1)} & = & \frac{i\hbar \alpha}{3\tau }\left[\bm{n}(\hat{\bm{j_1}}\times \hat{\bm{\sigma }})+\bm{n}(\hat{\bm{\sigma }}\times \hat{\bm{j_1}})\right]\label{A342}  \, , \\ 
\hat \Omega _{\text so}^{(2)} & = & \frac{2\hbar \alpha ^2}{3\tau }\left[\bm{\mu _{\text s}}\hat{\bm{\sigma }}+
(\bm{n}\hat{\bm{\sigma }})(\bm{n}\bm{\mu _{\text s}})+\hat{\bm{j_1}}\bm{n}\right]\label{A341} \, .
\end{eqnarray}
Similarly, for magnetic impurity scattering, from (\ref{A224}) we find 
\begin{equation}
\hat \Sigma _{\text sm}^R=-\frac{i\hbar }{2\tau _{\text sm}}\hat 1,\quad \hat \Sigma _{\text sm}^A=\frac{i\hbar }{2\tau _{\text sm}}\hat 1 \, , \nonumber
\end{equation}
\begin{equation}
\hat \Sigma _{\text sm}^K=-\frac{i\hbar }{2\tau _{\text sm}}\left(\mu _c\hat 1-\bm{\mu _{\text s}}\hat{\bm{\sigma }}\right) \, , \nonumber
\end{equation}%
and 
\begin{equation}\label{A355}
\hat \Omega _{\text sm}=\frac{\hbar }{\tau _{\text sm}}\bm{\mu _{\text s}}\hat{\bm{\sigma }}+\frac{\hbar }{2\tau _{\text sm}}\bm{j}_1\bm{n} \, .
\end{equation}
After substitution of (\ref{A290}), (\ref{A341}), (\ref{A342}), (\ref{A355}) into Eilenberger equation (\ref{A280}) and 
averaging over $d\bm{n}$ we find 
\begin{equation}\label{A360}
\frac{1}{3}v_F\bm{\nabla }\hat{\bm{j_1}}+\left(\frac{8\alpha ^2}{9\tau }+\frac{1}{\tau _{\text sm}}\right)\bm{\mu _{\text s}}\hat{\bm{\sigma }}=0 \, .
\end{equation}
Next, we find a second equation by averaging over $d\bm{n}$ the (\ref{A280}) multiplated by $\bm{n}$
\begin{eqnarray}\label{A370}
& v_F\tau \bm{\nabla }{\hat \mu }_o+(1+\frac{2\alpha ^2}{3})\hat{\bm{j_1}}+{}\nonumber\\
& \frac{i\alpha }{3}\left[(\hat{\bm{j_1}}\times \hat{\bm{\sigma }})+(\hat{\bm{\sigma }}\times \hat{\bm{j_1}})\right]=0 &  \, .
\end{eqnarray}
Eq. (\ref{A370}) can be solved to give 
\begin{eqnarray}\label{A380}
& \hat{\bm{j_1}}=-v_F\tau [\frac{1}{1+2\alpha ^2/3}\bm{\nabla }\hat \mu _o+K_1\bm{\nabla }(\bm{\mu _{\text s}}\hat{\bm{\sigma }})-K_2(\hat{\bm{\sigma }}\bm{\nabla })\bm{\mu _{\text s}}+{}\nonumber\\
& (K_2-K_1)\hat{\bm{\sigma }}(\bm{\nabla \mu _{\text s}})] \, , &
\end{eqnarray}
where
\begin{eqnarray}
K_1 & = & \frac{4\alpha ^2}{(3+2\alpha ^2)^2-4\alpha ^2} \, , \\
K_2 & = & \frac{2\alpha (3+2\alpha ^2)}{(3+2\alpha ^2)^2-4\alpha ^2} \, .
\end{eqnarray}
We now use the representation (\ref{A240}) to find from (\ref{A100}) the current caused by the normal velocity operator
\begin{equation}\label{A390}
\hat{\bm{j}_{\text{N}}}(\bm{r})=\frac{eN_o}{2}\int \frac{d\bm{n}}{4\pi }v_F\bm{n}\,({\hat \mu }_o+\hat{\bm{j_1}}\bm{n})=\frac{eN_ov_F}{6}\hat{\bm{j_1}} \, .
\end{equation}
After substitution (\ref{A380}) into equation above we find finally that there are three contributions to the current density from the normal velocity operator:
$$
\hat{\bm{j}}_{\text{N}}(\bm{r})=\hat{\bm{j}}_{\text{o}}(\bm{r})+\hat{\bm{j}}_{\text{sj}}(\bm{r})+\hat{\bm{j}}_{\text{ss}}(\bm{r}) \, 
$$
where
\begin{eqnarray}
\hat{\bm{j}}_{\text{o}} & = & -\frac{eN_oD}{2}[\frac{1}{1+2\alpha ^2/3}\bm{\nabla }\mu _c+ {}\nonumber\\
& &\frac{27+36\alpha ^2+20\alpha ^4}{3(3+2\alpha ^2)(1+2\alpha ^2)(3-2\alpha ^2)}\bm{\nabla }(\bm{\mu _{\text s}}\hat{\bm{\sigma }})] \label{A401} \\
\hat{\bm{j}}_{\text{sj}} & = & \frac{\alpha (3+2\alpha ^2)eN_oD}{(3+2\alpha ^2)^2-4\alpha ^2}(\hat{\bm{\sigma }}\bm{\nabla })\bm{\mu _{\text s}} \label{A402} \\
\hat{\bm{j}}_{\text{ss}} & = & -\frac{\alpha eN_oD}{2\alpha ^2+2\alpha +1}\hat{\bm{\sigma }}(\bm{\nabla \mu _{\text s}})  \label{A403}
\end{eqnarray}
Also using expression for $\hat{\bm{j_1}}$ (\ref{A380}) we can rewrite the formula for anomalous current (\ref{A260}) like
\begin{widetext}
\begin{eqnarray}\label{A410}
& \hat {\bm{j}}_{\text{av}}(\bm{r})=-\frac{eN_ov_F}{4k_F}\{\left(\frac{2\alpha ^2}{2\alpha ^2+2\alpha +1}\bm{\nabla }\times \bm{\bm{\mu _{\text s}}}
+\hat{\bm{\sigma }}\times \bm{\nabla }\mu _c\right)f_1(\alpha )+ {}\nonumber\\
& \sum_{i}\bm{e_i}\sum_{jk}\epsilon _{ijk}[\left(K_1(\alpha )f_2(\alpha )-K_2(\alpha )f_3(\alpha )\right)\nabla _k\mu _{s,j}+
\left(K_1(\alpha )f_3(\alpha )-K_2(\alpha )f_2(\alpha )\right)\nabla _j\mu _{s,k}]\} &
\end{eqnarray}
\end{widetext}
Diffusion equations for both spin and charge distribution functions can be derived directly
from the equation (\ref{A360}) after substitution (\ref{A380}):
\begin{equation}\label{A420}
\bm{\nabla ^2}\mu _c=0
\end{equation}
\begin{equation}\label{A430}
\frac{(2\alpha ^2+3)^2D}{(2\alpha ^2+3)^2-4\alpha ^2}\bm{\nabla ^2\mu _{\text s}}=\frac{\bm \mu _{\text s}}{\tau _{so}}+\frac{4\alpha ^2D}{(2\alpha ^2+3)^2-4\alpha ^2}\bm{\nabla }(\bm{\nabla }\bm{\mu _{\text s}}).
\end{equation}
In the limit of weak spin-orbit interaction ($\alpha \ll 1$) we obtain the simplified diffusion equations (\ref{II-30}) and (\ref{II-50}) and the simplified expression for the current density (\ref{II-60}), (\ref{II-80}), (\ref{II-70}), (\ref{II-90}), and (\ref{II-100}).

\section{CONCLUSIONS} \label{con}

We have derived diffusion equations for spin and charge flow in normal metals and the associated expression for the spin and charge currents. The total current consists of four contributions: Ordinary current, anomalous current, side-jump current and skew scattering current. These macroscopic diffusion equations, allows computation of charge Hall voltages and spin Hall voltages in pure normal metals or hybrid ferromagnet-normal metal systems.

We have considered two experimental relevant geometries and calculated Hall and spin-Hall voltages in the case of weak spin-orbit
interaction. In pure normal metals with no ferromagnetic contacts, there is no charge Hall effect due to the spin-orbit interaction, and the spin Hall effect is caused by the anomalous current, in agreement with the observation in Ref. \onlinecite{zhang}. In this geometry, we have made the additional observation that the spin Hall voltage reaches and "universal" value independent of the strength of the spin-orbit interaction, when spin-flip scattering due to spin-orbit scattering dominates spin-flip scattering due to magnetic impurities. When a spin-polarized current is injected into a normal metal film, both a Hall voltage and a spin Hall voltage exist. The magnitude of the Hall voltage is governed by side-jump, skew scattering and anomalous currents when the system is shorter than the spin-diffusion length. For systems longer than the spin-diffusion length, the Hall voltage vanishes, and the spin Hall effect resumes the value dominated by the anomalous current. In the intermediate regime, the competition between skew scattering, side-jump scattering and anomalous velocity operator determines the spin Hall and charge Hall voltages.

\begin{acknowledgements}
We would like to thank Andy Kent, Jan Petter Morten and Oleg Jouravlev for stimulating discussions. This work has been supported in part by the Research Council of Norway, NANOMAT Grants No. 158518/143 and 158547/431, and through Grant No. 153458/432.
\end{acknowledgements}

\end{document}